\documentclass[10pt,aps,prl,floatfix,showpacs,twocolumn,longbibliography,superscriptaddress,nofootinbib]{revtex4-1}
\usepackage[normalem]{ulem}
\usepackage{dcolumn}
\usepackage{bm}
\usepackage{amssymb}
\usepackage{amsmath}
\usepackage{graphicx}
\usepackage{amsfonts}
\usepackage{slashed}
\usepackage[colorlinks=true, linkcolor=blue, citecolor=blue, filecolor=blue,urlcolor=blue]{hyperref}
\usepackage[dvipsnames]{xcolor}
\usepackage{array}
\usepackage{multirow}
\allowdisplaybreaks
\usepackage{epsf}
\usepackage[nice]{nicefrac}
\usepackage{MnSymbol}
\usepackage{orcidlink}
\usepackage{physics}
\usepackage{tikz}
\usetikzlibrary{matrix,
fit,
calc,
backgrounds,
positioning,
arrows.meta,
decorations.pathreplacing,
shapes.arrows
}
\usepackage{float}

\newcommand{\btheta}{\boldsymbol{\theta}}
\begin{document}

\title{Neural Quantum States in Non-Stabilizer Regimes: Benchmarks with Atomic Nuclei}

\author{James W. T. Keeble\,\orcidlink{0000-0002-6248-929X}}
\email{jkeeble@physik.uni-bielefeld.de}
\affiliation{Fakult\"at f\"ur Physik, Universit\"at Bielefeld, D-33615, Bielefeld, Germany}

\author{Alessandro Lovato\,\orcidlink{0000-0002-2194-4954}}
\email{lovato@anl.gov}
\affiliation{Physics Division, Argonne National Laboratory, Argonne, Illinois 60439, USA}
\affiliation{Computational Science (CPS) Division, Argonne National Laboratory, Argonne, Illinois 60439, USA}
\affiliation{INFN-TIFPA Trento Institute for Fundamental Physics and Applications, Trento, Italy}

\author{Caroline E. P. Robin\,\orcidlink{0000-0001-5487-270X}}
\email{crobin@physik.uni-bielefeld.de}
\affiliation{Fakult\"at f\"ur Physik, Universit\"at Bielefeld, D-33615, Bielefeld, Germany}
\affiliation{GSI Helmholtzzentrum f\"ur Schwerionenforschung, Planckstra{\ss}e 1, 64291 Darmstadt, Germany}

\date{\today}

\begin{abstract}
As neural networks are known to efficiently represent classes of tensor-network states as well as volume-law-entangled states, identifying which properties determine the representational capabilities of neural quantum states (NQS) remains an open question. We construct NQS representations of ground states of medium-mass atomic nuclei, which typically exhibit significant entanglement and non-stabilizerness, to study their performance in relation to the quantum complexity of the target state. Leveraging a second-quantized formulation of NQS tailored for nuclear-physics applications, we perform calculations in active orbital spaces using a restricted Boltzmann machine (RBM), a prototypical NQS ansatz. For a fixed number of configurations, we find that states with larger non-stabilizerness are systematically harder to learn, as evidenced by reduced accuracy. This finding suggests that non-stabilizerness is a primary factor governing the compression and representational efficiency of RBMs in entangled regimes, and motivates extending these studies to more sophisticated network architectures.
\end{abstract}

\maketitle

\textit{Introduction.}---
Ongoing efforts towards solving quantum many-body (QMB) problems, together with the development of quantum information and quantum computation, have led to enormous progress in our understanding of quantum complexity in many-body systems, in connection with information theory~\cite{Baiguera:2025dkc}.
Entanglement is one long-recognized aspect of quantum complexity that contributes to the exponential growth of information required to represent quantum states, and is the bottleneck of many classical many-body methods, such as tensor-network techniques~\cite{Schollwoeck:2010uqf,Orus:2013kga,Banuls:2022vxp}.
While necessary, entanglement is however not sufficient to render a quantum state truly complex, as some highly-entangled stabilizer states may be prepared with polynomial classical resources~\cite{Gottesman:1998hu,Aaronson:2004xuh}, owing to the
regularity of their entanglement patterns such as flat entanglement spectra~\cite{Tirrito:2023fnw}. Non-stabilizerness (``quantum magic''~\cite{Bravyi:2004isx}) is thus a required ingredient to generate genuine quantum complexity, such as unstructured entanglement patterns, and correct multi-partite entanglement features~\cite{Hayden:2021gno,Cao:2024nrx}, leading to exponential scalings in quantum-state representation.
Recently-developed measures~\cite{Leone:2021rzd} have enabled the 
quantification of non-stabilizerness in various QMB systems, see {\it e.g.} Refs.~\cite{Oliviero:2022euv,Haug:2022vpg,Rattacaso:2023kzm,Tarabunga:2023ggd,Gu:2023ylw,Cao:2024nrx,Passarelli:2024tyi,Frau:2024qmf,Falcao:2024msg,Brokemeier:2024lhq,Hartse:2024qrv,Chernyshev:2024pqy,Sinibaldi:2025cst,Viscardi:2025vya,Sarkis:2025oab,Robin:2025wip,Santra:2025pvn,Santra:2025dsm,Grieninger:2026bdq}, providing indirect estimates of the computational complexity
in frameworks where entanglement can be produced at low cost, most commonly, quantum computations based on Clifford plus non-stabilizer resource theory~\footnote{In such resource theory Clifford operations can create highly-entangled stabilizer states, while non-stabilizerness, needed for 
universality, is the expensive resource.}.

At the same time, neural networks have appeared as powerful and versatile classical tools for describing QMB systems~\cite{Carleo:2016svm}. One appealing characteristic of these neural quantum states (NQS) is their ability to efficiently capture volume-law entanglement, even in their simplest Restricted Boltzmann Machine (RBM) form~\cite{Deng:2017uik,Gao:2017xhj}.
What physical property imposes a limitation on the representational capabilities of neural networks is thus still unclear, and studies exploring this question through connections with many-body complexity have developed, largely focusing on aspects of entanglement or classical information~\cite{Levine:2019gtf,Passetti:2022ilw,Denis:2023dww,Yang:2024yxu,Dash:2024wtl,Jreissaty:2025qip,Paul:2025uew,Lu:2026pgr}. Understanding the role of non-stabilizerness in NQS is a natural further step towards addressing this question,
and first investigations have been carried out for the Heisenberg model and random RBMs~\cite{Sinibaldi:2025cst}.

Nuclear many-body systems appear as ideal candidates to investigate the representation power of NQS, in connection with many-body quantum complexity. Due to the non-perturbative character of the strong nuclear force combined with the presence of two fermionic species (protons and neutrons), these systems present an extremely rapid growth of Hilbert space dimensionality, and have been found to
exhibit high degrees of complexity in their ground states, characterized by
volume-law entanglement entropies~\cite{Gu:2023aoc}, as well as significant
multipartite proton-neutron entanglement and non-stabilizerness~\cite{Brokemeier:2024lhq}.

Previous descriptions of nuclear systems with NQS have so far employed first-quantization frameworks, in which fermion antisymmetry must be encoded in the variational ansatz~\cite{Adams:2020aax,Lovato:2022tjh,Yang:2022esu,Yang:2023rlw,Gnech:2023prs,Yang:2025mhg}. To the best of our knowledge, only a study of the pairing model has applied second-quantization techniques~\cite{Rigo:2022ces}.
In the present work, we generalize second-quantization NQS methods originally developed in quantum chemistry and condensed-matter physics~\cite{Choo:2019vcr} to represent ground states of atomic nuclei. Specifically, we employ a RBM ansatz defined over occupation-number configurations and perform calculations within an active orbital space using a nucleon–nucleon interaction known to successfully reproduce low-energy experimental nuclear data. Exact diagonalization within that space is referred to as the interacting shell model (ISM)~\cite{Brown:2001zz,Caurier:2004gf,Otsuka:2018bqq}, and is analogous to complete active-space configuration interaction in quantum chemistry. Despite the substantial reduction in dimensionality compared to no-core \textit{ab initio} approaches~\cite{Barrett:2013nh}, the ISM still exhibits 
a similar
combinatorial growth of the configuration basis, making it unfeasible for heavy nuclei that require large active spaces, even on leadership-class supercomputers.
As the ISM has recently been used to study measures of many-body quantum complexity in nuclei~\cite{Papenbrock:2003bj,Johnson:2022mzk,Perez-Obiol:2023wdz,Brokemeier:2024lhq,Shinde:2025xud}, it provides an ideal testbed for investigating whether NQS can capture the relevant nuclear correlations without incurring the exponential scaling of exact diagonalization.

We apply the second-quantized NQS framework to a range of medium-mass nuclei in the $sd$-shell. Comparisons with ISM calculations allow us to quantify the performance of the neural representation, and we show that the attainable accuracy correlates with measures of non-stabilizerness of the ground states, thereby establishing a direct connection between representational complexity and intrinsic quantum complexity in nuclear wave functions.
\\

\textit{Method.}--- 
The 
nucleon orbitals are partitioned into a fully filled core and a partially occupied active space, taken to be one harmonic oscillator shell, in which nucleons interact. Specifically, we work in the $sd$-shell within the so-called $m$-scheme, where orbitals are characterized by $a \equiv (n_a,l_a,j_a,m_a=j_{z_a}, \tau_a)$ quantum numbers (principal quantum number, orbital and total angular momenta, projection of the total angular momentum, and isospin projection). The $sd$-shell then comprises $N_\pi = 12$ proton and $N_\nu = 12$ neutron orbitals, which generate wave-functions with up to $\sim 10^5$ many-body configurations within a given symmetry sector defined by proton number, neutron number, and projection of total angular momentum $J_z$. 
We employ the high-quality \textit{USDB} parametrization of the two-body nuclear Hamiltonian~\cite{PhysRevC.74.034315} 
\begin{equation}
    \hat H = \sum_a \varepsilon_a c^\dagger_a c_a + \frac{1}{4} \sum_{abcd} \tilde{v}_{ab,cd} \, c^\dagger_a c^\dagger_{b} c_{d} c_{c} \; ,
\end{equation}
where $c^\dagger_a$ ($c_a$) creates (annihilates) a nucleon in state $a$, $\varepsilon_a$ are the single-nucleon energies, and $\tilde{v}_{ab,cd}$ denotes antisymmetrized interaction matrix elements.

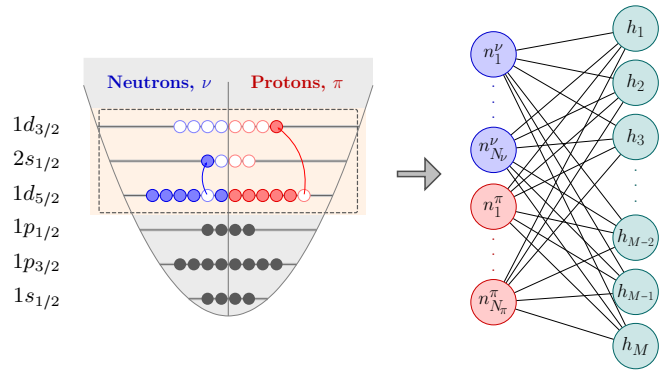
\begin{figure}[htbp]
\centering
\resizebox{\columnwidth}{!}{%

\begin{tikzpicture}[>=stealth]

    \def\labelx{4.8}
    \def\circlesize{5pt}

    \begin{scope}[scale=2, shift={(-6.5, 0)}]

        \fill[black!8, domain=-4.2:4.2, smooth, variable=\x]
            plot ({\x}, {0.38*\x*\x - 5.5}) -- (4.2, 2) -- (-4.2, 2) -- cycle;

        \node[blue!75!black, font=\Huge\bfseries, scale=1.2] 
            at (-2., 1.25) {Neutrons, $\nu$};
        \node[red!75!black, font=\Huge\bfseries, scale=1.2]
            at (2., 1.25) {Protons, $\pi$};

        \fill[orange!10] (-4.0,-2.55) rectangle (4.0,0.55);

        \foreach \y/\orblabel/\halfwidth in {
          -5/{$1s_{1/2}$}/1.147,
          -4/{$1p_{3/2}$}/1.987,
          -3/{$1p_{1/2}$}/2.565,
          -2/{$1d_{5/2}$}/3.035,
          -1/{$2s_{1/2}$}/3.441,
           0/{$1d_{3/2}$}/3.804
        } {
          \draw[black!20, line width=5pt]   (-\halfwidth,\y) -- (\halfwidth,\y);
          \draw[black!60, line width=1.6pt] (-\halfwidth,\y) -- (\halfwidth,\y);
          \node[left, black!100] at (-\labelx,\y)
              {\scalebox{4}{\bfseries\itshape\orblabel}};
        }

        \foreach \x in {-0.6, -0.2}
            { \fill[gray!70!black] (\x,-5) circle (\circlesize); }
        \foreach \x in {-1.4, -1.0, -0.6, -0.2}
            { \fill[gray!70!black] (\x,-4) circle (\circlesize); }
        \foreach \x in {-0.6, -0.2}
            { \fill[gray!70!black] (\x,-3) circle (\circlesize); }

        \foreach \x in {-0.6, -0.2}
            { \fill[gray!70!black] (-\x,-5) circle (\circlesize); }
        \foreach \x in {-1.4, -1.0, -0.6, -0.2}
            { \fill[gray!70!black] (-\x,-4) circle (\circlesize); }
        \foreach \x in {-0.6, -0.2}
            { \fill[gray!70!black] (-\x,-3) circle (\circlesize); }

        \foreach \x in {-2.2, -1.8, -1.4, -1.0, -0.2}
            { \fill[blue!50] (\x,-2) circle (\circlesize);
              \draw[blue!90!black, line width=0.5pt] (\x,-2) circle (\circlesize); }

        \fill[white] (-0.6,-2) circle (\circlesize);
        \draw[blue!50, line width=1.6pt] (-0.6,-2) circle (\circlesize);

        \fill[blue!50] (-0.6,-1) circle (\circlesize);
        \draw[blue!90!black, line width=0.5pt] (-0.6,-1) circle (\circlesize);

        \fill[white] (-0.2,-1) circle (\circlesize);
        \draw[blue!50, line width=1.6pt] (-0.2,-1) circle (\circlesize);

        \foreach \x in {-0.2, -0.6, -1.0, -1.4}
            { \fill[white] (\x,0) circle (\circlesize);
              \draw[blue!50, line width=1.6pt] (\x,0) circle (\circlesize); }

        \foreach \x in {-1.8, -1.4, -1.0, -0.6, -0.2}
            { \fill[red!50]  (-\x,-2) circle (\circlesize);
              \draw[red!90!black, line width=0.5pt] (-\x,-2) circle (\circlesize); }
        
        \fill[white] (2.2,-2) circle (\circlesize);
        \draw[red!50, line width=1.6pt] (2.2,-2) circle (\circlesize);

        \fill[red!50] (1.4,0) circle (\circlesize);
        \draw[red!90!black, line width=0.5pt] (1.4,0) circle (\circlesize);

        \foreach \x in {0.2, 0.6}
            { \fill[white] (\x,-1) circle (\circlesize);
              \draw[red!50, line width=1.6pt] (\x,-1) circle (\circlesize); }

        \foreach \x in {0.2, 0.6, 1.0}
            { \fill[white] (\x,0) circle (\circlesize);
              \draw[red!50, line width=1.6pt] (\x,0) circle (\circlesize); }

        \draw[black!65, line width=0.7mm, rounded corners=3pt,
              dash pattern=on 7pt off 3pt]
            (-3.75,-2.5) rectangle (3.75,0.5);

        \draw[->, blue!100!black, line width=1.6pt, bend left=30,
              shorten >=2pt, shorten <=2pt]
            (-0.6,-2) to (-0.6,-1);
        \draw[->, red!100!black, line width=1.6pt, bend right=30,
              shorten >=2pt, shorten <=2pt]
            (2.2,-2) to (1.4, 0);

        \draw[=>, thick, black!55] (0,-5.5) -- (0,1.3);

        \draw[thick, black!50, domain=-4.2:4.2, smooth, variable=\x]
            plot ({\x}, {0.38*\x*\x - 5.5});

    \end{scope}

    \node[single arrow,
          draw=black!70, fill=black!25,
          minimum height=2.5cm, minimum width=1.8cm,
          single arrow head extend=0.75cm,
          line width=2.5pt]
        at (-2, -2.75) {};
    
    \begin{scope}[scale=1.2, shift={(2, 3.5)}, x=2.5cm, y=2.5cm]

        \tikzstyle{visible proton}=[
            circle, draw=red!80!black, line width=1.2pt,
            fill=red!20, minimum size=75pt, inner sep=0pt,
            font=\Huge, text=black!90!black]
        \tikzstyle{visible neutron}=[
            circle, draw=blue!80!black, line width=1.2pt,
            fill=blue!20, minimum size=75pt, inner sep=0pt,
            font=\Huge, text=black!90!black]
        \tikzstyle{hidden neuron}=[
            circle, draw=teal!80!black, line width=1.2pt,
            fill=teal!20, text=black!90!black, minimum size=75pt,
            inner sep=0pt, font=\Huge]

        \node[visible neutron] (vnu1) at (0,  0.000) {\scalebox{1.25}{$n^{\nu}_{1}$}};
        \node[font=\Huge, color=blue!70!black] at (0, -0.65) {$\cdot$};
        \node[font=\Huge, color=blue!70!black] at (0, -0.95) {$\cdot$};
        \node[font=\Huge, color=blue!70!black] at (0, -1.25) {$\cdot$};
        \node[visible neutron] (vnuN) at (0, -1.85) {\scalebox{1.25}{$n^{\nu}_{N_{\!\nu}}$}};

        \node[visible proton]  (vpi1) at (0, -2.95) {\scalebox{1.25}{$n^{\pi}_{1}$}};
        \node[font=\Huge, color=red!70!black] at (0, -3.60) {$\cdot$};
        \node[font=\Huge, color=red!70!black] at (0, -3.90) {$\cdot$};
        \node[font=\Huge, color=red!70!black] at (0, -4.20) {$\cdot$};
        \node[visible proton]  (vpiN) at (0, -4.80) {\scalebox{1.25}{$n^{\pi}_{N_{\!\pi}}$}};

        \node[hidden neuron] (h1) at (2.75,  0.50) {\scalebox{1.25}{$h_{1}$}};
        \node[hidden neuron] (h2) at (2.75, -0.55) {\scalebox{1.25}{$h_{2}$}};
        \node[hidden neuron] (h3) at (2.75, -1.60) {\scalebox{1.25}{$h_{3}$}};
        \node[font=\Huge, color=teal!70!black] at (2.75, -2.27) {$\cdot$};
        \node[font=\Huge, color=teal!70!black] at (2.75, -2.57) {$\cdot$};
        \node[font=\Huge, color=teal!70!black] at (2.75, -2.87) {$\cdot$};
        \node[hidden neuron] (h4) at (2.75, -3.55)
        {\scalebox{1.25}{$h$}$_{M-2}$};
        \node[hidden neuron] (h5) at (2.75, -4.60)
        {\scalebox{1.25}{$h$}$_{M-1}$};
        \node[hidden neuron] (hM) at (2.75, -5.65) {\scalebox{1.25}{$h_{M}$}};

        \foreach \vi in {vpi1, vpiN, vnu1, vnuN} {
            \foreach \hj in {h1, h2, h3, h4, h5, hM} {
                \draw[black!100, line width=1pt] (\vi) -- (\hj);
            }
        }

    \end{scope}

\end{tikzpicture}
}%
\caption{Mapping the ISM to the RBM ansatz: active proton and neutron orbitals are mapped onto the $N = N_\nu + N_\pi$ visible nodes which carry the occupation numbers $n_i$ of these orbitals, and are correlated via their connections to the hidden nodes. 
}
\label{fig:NQS_ISM_mapping}
\end{figure}

Our analysis is based on a relatively simple RBM ansatz, which has proven accurate in describing molecules in the occupation-number formalism~\cite{Choo:2019vcr}. The RBM can be viewed as a neural network comprising a single fully connected layer with $N$ visible nodes and $M$ hidden nodes, with hidden-unit density $\alpha = M/N$. It defines a variational map between the discrete visible degrees of freedom and hidden variables in a space whose dimension is equal to or larger than that of the visible layer ($\alpha \geq 1$). The mapping of nuclear quantum state onto the network is illustrated in Fig.~\ref{fig:NQS_ISM_mapping}.

\noindent The nuclear NQS is then given by
\begin{align}
    \ket{ \Psi_{\btheta} } = \sum_{n} \langle n | \Psi_{\btheta} \rangle \ket{n} \; ,
\end{align}
where $\btheta$ denotes the variational network parameters, and $n$ takes integer values that specify the many-nucleon configurations (computational-basis states)
\begin{align}
n \equiv (n^\pi_1,  ... n^\pi_{N_\pi}, n^\nu_1,  ... n^\nu_{N_\nu} ) \equiv (n_1, .... n_N) \; ,
\end{align}
with 
$n_i$ taking values $-1$ or $+1$, corresponding to occupation numbers $0$ or $1$ respectively, while 
satisfying global symmetry constraints.
The RBM wave function amplitude takes the form~\cite{Carleo:2016svm}
\begin{equation}
    \braket{ n}{ \Psi_{\btheta} } = \sum_{h_i} \exp \left( \sum_j a_j n_j + \sum_i b_i h_i + \sum_{ij} \mathcal{W}_{ij} h_i n_j \right) \; ,
\end{equation}
where the parameters $\btheta = \{\mathcal{W}_{ij}, b_i, a_j\}$ are complex-valued and scale quadratically in the number of visible nodes, {\it i.e.} as $\mathcal{O}(N \cdot M) = \mathcal{O}(\alpha N^2)$.
Tracing out over the hidden values $h_i \in \{-1, +1\}$ one can further simplify the amplitudes as~\cite{Carleo:2016svm}
\begin{equation}
    \ln(\braket{ n }{ \Psi_{\btheta} }) = \sum_{i=1}^N a_i n_i + \sum_{j=1}^M \ln(2\cosh(b_j + \sum_{i=1}^N \mathcal{W}_{ji} n_i)) \; .
\end{equation}
\\

In order to assess the representation capabilities of the network, we first optimize the set of parameters $\btheta$ by maximizing the fidelity $\mathcal{F}_{\btheta}$ of the NQS state with respect to the exact (normalized) ISM ground state $\ket{\Psi_{ex}}$ obtained via full diagonalization in the configuration space
\begin{equation}
    \mathcal{F}_{\btheta} = \frac{|\langle \Psi_{\btheta}| \Psi_{ex}\rangle |^2}{\langle \Psi_{\btheta} | \Psi_{\btheta} \rangle  } \; .
\end{equation}
As a second step, towards using NQS for predicting quantum states in heavier nuclei where exact diagonalization and full-summation over the basis states are not possible, we optimize $\btheta$ through minimization of the NQS energy
\begin{equation}
    E_{\btheta} = \frac{\langle \Psi_{\btheta}| \hat H |\Psi_{\btheta}\rangle }{\langle \Psi_{\btheta} | \Psi_{\btheta} \rangle  } \; ,
\end{equation}
via variational Monte Carlo (VMC) sampling techniques. 
For both procedures, the numerical optimizations are performed using the RMSProp variant of the Stochastic Reconfiguration (SR) algorithm~\cite{Sorella:2005}, which has a wide breadth of successful application of optimizing NQSs over a variety of nuclear systems~\cite{Lovato:2022tjh,Fore:2024exa,DiDonna:2025oqf}. We provide a brief summary of the method in the end matter.
\\

In this work we investigate the performance of the NQS depending on the intrinsic quantum complexity of the nuclear state. To characterize this complexity
we consider measures of non-stabilizerness quantified by stabilizer R\'enyi entropies (SREs)~\cite{Leone:2021rzd}.
Specifically, we consider the 2-R\'enyi entropy which possesses the correct property of monotonicity~\cite{Leone:2024lfr,Haug:2023hcs} and is defined as
\begin{equation}
    \mathcal{M}_2 (\ket{\Psi}) =  - \log_2 \left( \sum_P \frac{\langle \Psi | \hat P | \Psi \rangle^{4}}{2^N} \right) \; ,
\end{equation}
where $\hat P$ denotes tensor products of Pauli operators with overall phase $+1$.
In some sense the SREs measure the deviation between the distribution of $\ket{\Psi}$ onto Pauli strings, and the regular distribution of stabilizer states. In particular, $\mathcal{M}_2$
relates to the distance to the nearest stabilizer state~\cite{Haug:2024ptu}. The values of $\mathcal{M}_2$ in $sd$-shell nuclear ground states were computed in Ref.~\cite{Brokemeier:2024lhq} using the same orbital-to-node mapping as in the present work.
\\

\textit{Fidelity maximization.}--- 
\begin{figure*}[t]
\centering
  \includegraphics[width=\textwidth]{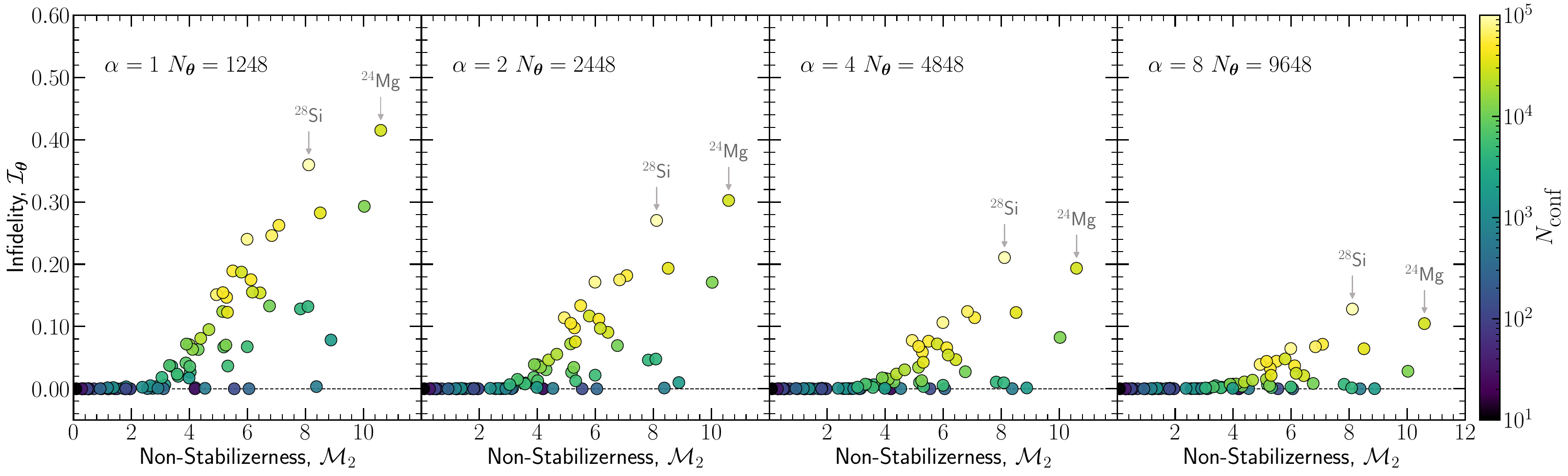}
\caption{
Infidelity $\mathcal{I}_{\btheta} = 1- \mathcal{F}_{\btheta}$ of the nuclear NQS obtained from fidelity maximization, as a function of non-stabilizerness $\mathcal{M}_2(\ket{\Psi_{ex}})$ in the exact ISM state, for different values of $\alpha$. The color code denotes the number of configurations $N_\mathrm{conf}$ in the chosen symmetry sector. The number of parameters $N_{\btheta}$ in the NQS is also indicated for each $\alpha$.}
\label{fig:fidmax_M2}
\end{figure*}   
Fig.~\ref{fig:fidmax_M2} shows the infidelity $\mathcal{I}_{\btheta} = 1- \mathcal{F}_{\btheta}$ of the fidelity-optimized NQS, as a function of the non-stabilizerness $\mathcal{M}_2(\ket{\Psi_{ex}})$ in the exact ISM ground state, for different values of the hidden-unit density $\alpha$. The colors represent the size of the many-body Hilbert space in the chosen symmetry sector, measured by the number of many-body configurations $N_{\mathrm{conf}}$, for each nucleus. Since the employed interaction is isospin-symmetric, only nuclei with neutron number greater or equal than proton number are displayed. The numerical values are shown in Table~\ref{tab:fidelity_full} of the supplemental material.

For $\alpha=1$, the network has $N_{\btheta} = 1248$ parameters and the NQS is therefore able to represent nuclear ground states with up to this many configurations with fidelities $\mathcal{F}_{\btheta} > 99.6\%$. Above this threshold, however, the fidelity of the nuclear NQS begins to worsen. While overall, we observe a global decrease in fidelity as the number of configurations $N_{\mathrm{conf}}$ increases, there is not a one-to-one correspondence between $N_{\mathrm{conf}}$ and the accuracy of the network.
In fact, it appears that, {\it for a fixed value of $N_{\mathrm{conf}} \gtrsim N_{\btheta}$, states exhibiting a higher degree of non-stabilizerness are systematically captured with lower fidelity}~\footnote{Some slight exceptions are nuclei around $^{32}$S which, interestingly, exhibits a large degree of non-local magic~\cite{Keeble2026_prep}.}. 
For example, $^{24}$Mg, with only 28,503 configurations, exhibits the largest amount of non-stabilizerness 
and is represented by the NQS with the lowest fidelity. Note that $^{24}$Mg is one of the most axially deformed nucleus in the $sd$-shell~\cite{nudat3}. 
As $\alpha$ increases, the number of parameters and expressivity of the network become larger and thus the attained accuracy of the NQS globally improves. However the correlation between non-stabilizerness and fidelity persists, for all $\alpha$ values. This indicates that the ability of the RBM to compress the nuclear states, and describe the relevant physics with fewer parameters, is limited by the non-stabilizer content of the state.

It is worth noting that no comparable correlations are observed between attained accuracy and simple measures of entanglement, such as single-orbital entanglement entropy.
In contrast correlations do appear 
for measures characterizing multi-partite entanglement in the mixed proton-neutron sector, but not in the pure proton or pure neutron sectors. 
This observation is consistent with the findings of Ref.~\cite{Brokemeier:2024lhq} which revealed correlated patterns for the multi-proton-neutron entanglement and non-stabilizerness in $sd$-shell nuclei.
More generally, it has been shown in different contexts that multi-partite entanglement features can be largely affected by the presence of (non-local) non-stabilizerness~\cite{Cao:2024nrx}.
We also note that the non-stabilizerness of the nuclear NQS is systematically predicted below the exact value, while this is not always the case for entanglement measures. These findings are documented in the end matter.

Overall, with $\alpha =8$, the wave functions of $sd$-shell nuclei are reproduced within less than $13 \%$ error in the worst cases, 
while we find that the relative errors on the energies are less than $3.5 \%$.
Note that for $^{28}$Si, the largest nucleus of the $sd$-shell, a network with $\alpha=8$ has a number of parameters which is only about $10 \%$ of the total number of configurations.
\\

\textit{Energy-minimization with VMC.}--- 
While the analysis above allows us to assess the expressivity and representational efficiency of NQS, the systems of ultimate interest are those for which exact solutions and full summations over basis states are not feasible. In such cases, one typically resorts to energy minimization using Monte Carlo sampling techniques. This requires rewriting the normalized Hamiltonian expectation value as an expectation over the local energy~\cite{Lovato:2026erx},
\begin{equation}
\frac{\langle \Psi |\hat H|\Psi \rangle}{\langle \Psi | \Psi\rangle}
= \mathbb{E}_{n \sim \pi_\Psi} \left[\frac{\bra{n}\hat{H}\ket{\Psi}}{\bra{n}\ket{\Psi}}\right] \; ,
\end{equation}
where the samples are drawn from $\pi_\Psi(n) = |\Psi(n)|^2 / \langle \Psi | \Psi\rangle$. We use the Metropolis--Hastings (MH) algorithm to generate physically valid configurations in the computational basis while preserving the relevant symmetries. 
Details of the sampling procedure are provided in the end matter.

\begin{figure*}[t]
\centering
  \includegraphics[width=\textwidth]{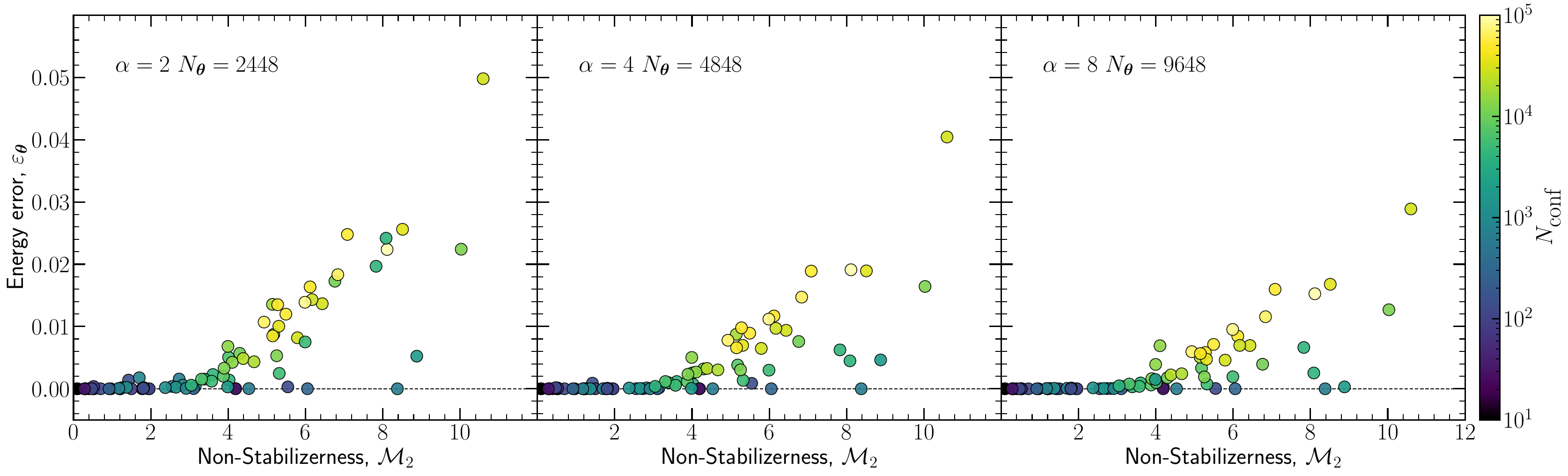}
  \includegraphics[width=\textwidth]{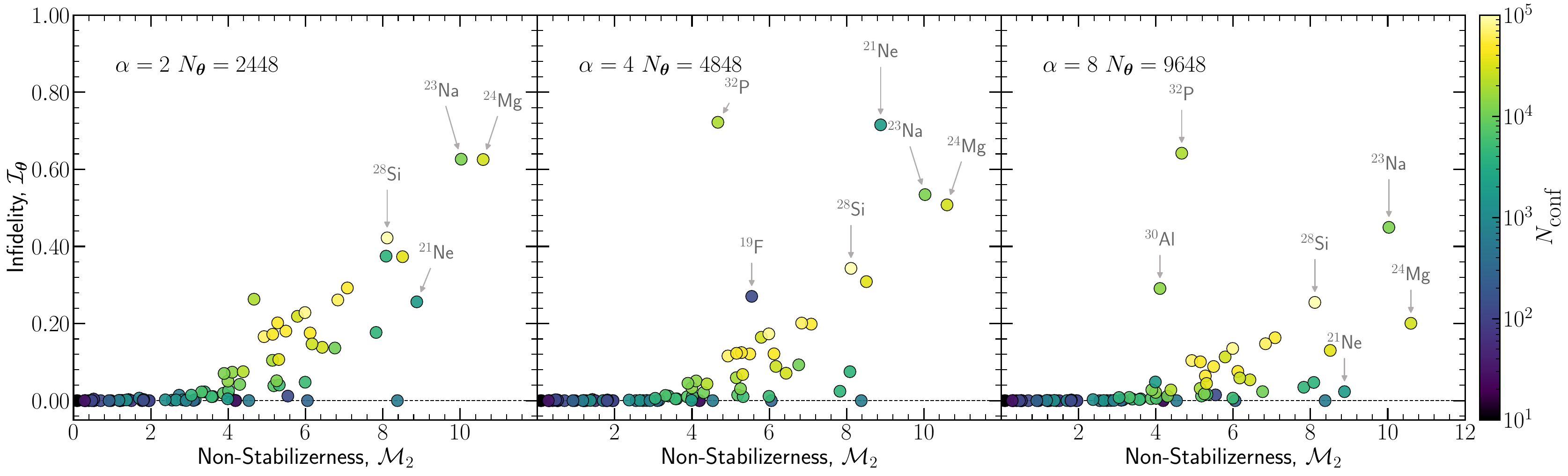}
\caption{Relative energy error $\varepsilon_{\btheta}$ and infidelity $\mathcal{I}_{\btheta}$ of the nuclear RBMs obtained via energy minimization with VMC.
}
\label{fig:VMC_M2}
\end{figure*}

Fig.~\ref{fig:VMC_M2} shows the relative energy error $\varepsilon_{\btheta} = |E_{ex}-E_{\btheta}|/|E_{ex}|$ and infidelity $\mathcal{I}_{\btheta} = 1- \mathcal{F}_{\btheta}$ of the resulting NQS. 
Except for a few problematic nuclei with convergence issues, both energies and wave functions show a similar correlation with the non-stabilizerness, as was observed with the fidelity-maximization procedure above, {\it i.e.} for states with $N_\mathrm{conf}> N_{\btheta}$ the accuracy reached by the network is typically lower for nuclei with large non-stabilizer content, while nuclei that are simpler from a quantum complexity point of view are easier to learn.

We observe that the energy is reproduced within $5\%$ accuracy for $\alpha = 2$ and within $3\%$ accuracy for $\alpha = 8$. Capturing the wave function itself, however, appears to be more challenging when the network is optimized solely through energy minimization. This behavior is well known and has been linked to the high dimensionality of variational spaces, which can give rise to flat energy landscapes; see, \textit{e.g.}, Ref.~\cite{Lange:2024nsr}. In addition, nuclei with odd numbers of protons and/or neutrons tend to be more difficult to converge to their exact ground states. Due to pairing effects, these typically exhibit significantly smaller gaps to their first excited states than even-even nuclei, and the optimization often converges to superpositions of ground and excited states. For example, $^{19}$F, $^{21}$Ne, $^{23}$Na, and $^{32}$P have excitation gaps between 77 and 399 keV, whereas even-even nuclei typically possess gaps of the order of $1.5$--$2$ MeV. Similar performance issues have recently been encountered in spin systems with small excitation gaps combined with a complicated sign or amplitude structure of the wave function~\cite{Cortes:2025htq}. The origin of this behavior is that the convergence rate of the SR algorithm depends strongly on the size of the gap. In practice, we find that, in some cases, adding a constraint to the Hamiltonian to enforce a fixed value of the total angular momentum $J$ improves convergence.

\textit{Summary and Discussion.}---
We have developed a NQS representation of nuclear ground states in the occupation-number basis and applied it to compute ground states of nuclei in the $sd$ shell. While such second-quantization techniques have been employed in quantum chemistry and condensed-matter physics, to the best of our knowledge this is the first time they are applied to atomic nuclei, systems which typically exhibit large-scale volume-law entanglement patterns. Our methodology is based on a RBM architecture with a varying number of hidden nodes, whose parameters are optimized using either fidelity maximization or VMC energy minimization. This framework has enabled the first NQS calculations of nuclei beyond light systems. 

We performed a systematic analysis of NQS accuracy in relation to the degree of non-stabilizerness of the quantum state, which, in the presence of large entanglement, governs the quantum complexity of the system. We found that ground states exhibiting large amounts of non-stabilizerness are more difficult for the network to learn, indicating reduced compressibility and representational efficiency within the RBM ansatz. This suggests that non-stabilizerness likely constitutes a bottleneck for the efficiency of RBMs in systems featuring volume-law entanglement, and serves as a key indicator of the network's efficiency and representational capacity in this regime. Our findings motivate extending this analysis to more advanced networks, including the recently developed Vision Transformer architecture~\cite{Viteritti:2022fji}. More broadly, we anticipate that understanding the underlying structure of quantum complexity for specific classes of problems will help guide the design of network architectures capable of efficiently capturing the targeted systems.

As measures of quantum complexity in fermionic systems are typically representation dependent, basis-optimization techniques aimed at reducing state complexity~\cite{Robin:2023pgi,Cortes:2025htq}, together with physically guided initial states such as stabilizer states~\cite{Zhang:2018jqb,Jia:2018vsu,Pei:2021yek,Sun:2024bvn,Robin:2025wip,Mao:2026bjq}, may accelerate convergence and help alleviate optimization issues encountered in the energy-minimization procedure. These aspects will be explored in future work.

\textit{Acknowledgments}---We would like to thank Giuseppe Carleo and Martin J. Savage for insightful discussions, and for their useful comments on the manuscript. We also thank Patrick Fasano, Gaute Hagen, and Mauro Rigo for valuable discussions.
The present research is supported, in part, by the U.S. Department of Energy, Office of Science, Office of Nuclear Physics, under contracts DE-AC02-06CH11357 (A.~L.), by the DOE Early Career Research Program (A.~L.), by the SciDAC-5 NeuCol program (A.~L).
The present work is also supported by the Ministerium f\"ur Kultur und Wissenschaft des Landes Nordrhein-Westfalen (MKW NRW) under the funding code NW21-024-A (J.~K. and C.~R.). 
The computations were performed, in part, on the GPU cluster at Bielefeld University. 
We thank the Bielefeld HPC.NRW team for their support. The authors also gratefully acknowledge the computer resources at Artemisa, funded by the European Union ERDF and Comunitat Valenciana as well as the technical support provided by the Instituto de Fisica Corpuscular, IFIC (CSIC-UV).

\bibliography{biblio_paper}

\onecolumngrid
\section*{End Matter}
\textit{Numerical Optimization.}--- 
Following common practice in neural-network quantum state applications across a range of physical systems~\cite{Lovato:2026erx}, in this work we employ the Stochastic Reconfiguration (SR) algorithm~\cite{Sorella:2005} to optimize the variational parameters. Within this framework, the parameter update at epoch $t$ is given by
\begin{align}
\btheta_{t+1} = \btheta_{t} - \eta G^{-1}_{t} \mathbf{g}_{t}\; ,
\end{align}
where $\eta$ denotes the learning rate. The quantum geometric tensor is defined as
\begin{equation}
G_{ij} = \frac{\langle \Psi_{\btheta}| O_i^\dagger O_j |\Psi_{\btheta}\rangle }{\langle \Psi_{\btheta} | \Psi_{\btheta} \rangle } - \frac{\langle \Psi_{\btheta}| O_i^\dagger |\Psi_{\btheta}\rangle }{\langle \Psi_{\btheta} | \Psi_{\btheta} \rangle } \frac{\langle \Psi_{\btheta}| O_j |\Psi_{\btheta}\rangle }{\langle \Psi_{\btheta} | \Psi_{\btheta} \rangle }\; ,
\end{equation}
and the gradient of the energy with respect to the variational parameters is given by
\begin{equation}
g_{i} = 2\left(\frac{\langle \Psi_{\btheta}| O_i^\dagger H |\Psi_{\btheta}\rangle }{\langle \Psi_{\btheta} | \Psi_{\btheta} \rangle } - \frac{\langle \Psi_{\btheta}| O_i^\dagger|\Psi_{\btheta}\rangle }{\langle \Psi_{\btheta} | \Psi_{\btheta} \rangle } \frac{\langle \Psi_{\btheta}| H |\Psi_{\btheta}\rangle }{\langle \Psi_{\btheta} | \Psi_{\btheta} \rangle }\right)\; ,
\end{equation}
where $O_i|\Psi_{\btheta}\rangle = \partial_{\theta_i} |\Psi_{\btheta}\rangle$ denotes the derivative of the wave function with respect to the $i$th variational parameter.

Although the QGT is, in principle, positive semi-definite, finite sampling effects may, in practice, generate negative eigenvalues. 
To remedy to this, the QGT is stabilized by adding a small positive constant to its diagonal component, $\lambda \sim 10^{-3}$.
However, as stated in Ref.~\cite{Lovato:2022tjh}, applying the same shift to all eigenvalues of the QGT washes out small eigenvalue information. Thus, we mitigate this numerical drawback by following the RMSprop-inspired regularization scheme of Ref.~\cite{Lovato:2022tjh}. 
This defines our regularization as,
\begin{align}
    G_t \rightarrow G_t + \lambda \, \textrm{diag}(\sqrt{\mathbf{v}_t} + \epsilon) \; ,
\end{align}
where,
\begin{align}
\mathbf{v}_t = \beta \mathbf{v}_{t-1} + (1 - \beta) \mathbf{v}^2_{t} \; ,
\end{align}
is the exponential moving average of the squared gradients with constant, $\beta$. The regularization parameter, $\epsilon$, is heuristically optimized at each epoch in order to obtain the ideal regularization parameter~\cite{Adams:2020aax}.
\\

\textit{Monte-Carlo Sampling.}--- 
In the VMC energy minimization procedure, a Metropolis-Hastings (MH) algorithm is used to generate configurations while preserving the relevant symmetries. We begin by selecting many-body configurations at random with the correct proton and neutron numbers. An initial MH procedure is then carried out in the space of total angular-momentum projection \(J_z\) to enforce the desired \(J_z\) value. After this setup, MH sampling is used to draw configurations from
\begin{equation}
\pi_\Psi(n) = \frac{|\Psi(n)|^2}{\langle \Psi | \Psi\rangle}\,.
\end{equation}
To traverse the configuration space efficiently while avoiding large rejection rates and ergodicity breaking, we define the proposal distribution in terms of single- and double-orbital swaps, chosen with probabilities of $90\%$ and $10\%$, respectively. 
These swaps are performed separately within the proton and neutron sectors. Cross-sector double swaps are also allowed to generate mixed proton-neutron two-nucleon excitations. 
Importantly, restricting the moves to swaps preserves the proton and neutron numbers in each MH update. This algorithm enables efficient generation of many-body configurations, albeit at the cost of autocorrelations in the evaluation of many-body observables. We mitigate autocorrelation by taking the number of sweeps between measurements of the local energy to be 10 times the number of single-particle orbitals, which in the case of the \emph{sd}-shell is 240. Such autocorrelations can be monitored by Gelman and Rubin's potential scale reduction of Refs.~\cite{gelman1992inference,Vehtari_2021,cabezas2024blackjax}. All statistical estimators are computed propagating in parallel 4096 chains and performing 10 MH steps for each, totaling 40,960 samples.
\\

\begin{figure}[h]
\centering
 \includegraphics[width=\textwidth]{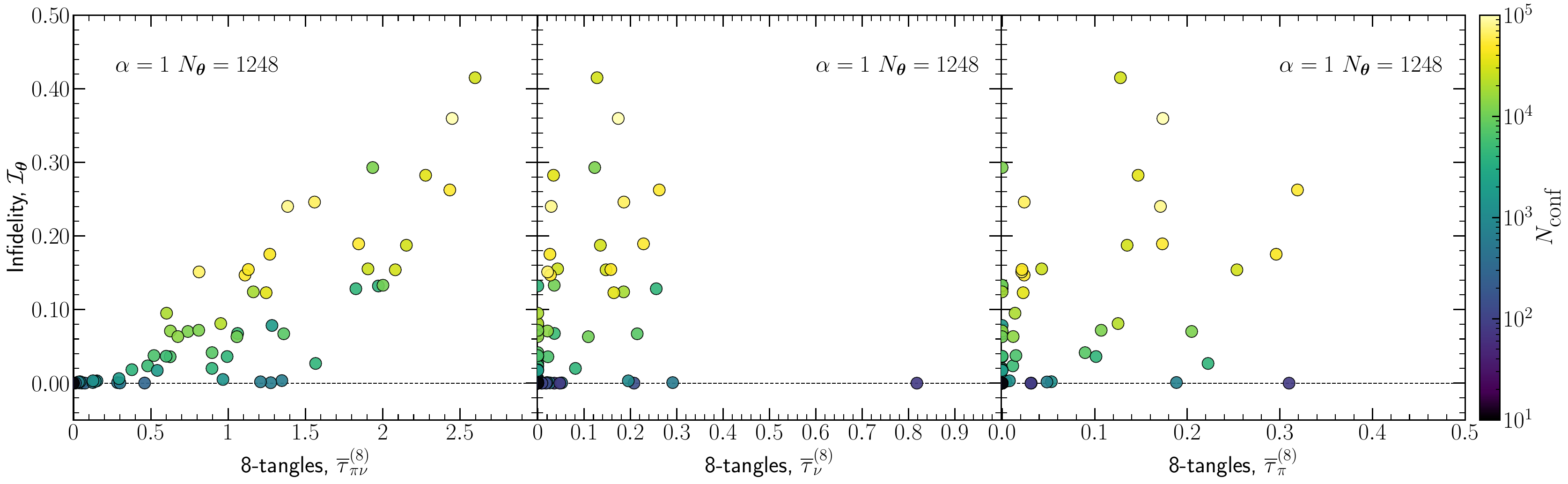}
\caption{Fidelity of the nuclear NQS versus multi-partite 8-orbital entanglement in the exact state, in the mixed proton-neutron (left), pure neutron (middle) and pure proton (right) sectors, for $\alpha=1$. }
\label{fig:tau8_fid}
\end{figure}

\textit{Multi-Partite Entanglement and NQS Accuracy.}--- 
Here we present how the accuracy reached by the NQS behaves depending on the 
multi-orbital entanglement properties of the exact ISM state. 
As a measure we choose the $8$-tangles~\cite{PhysRevA.63.044301}, defined as 
\begin{align}
    \tau^{(8)}_{(i_1 ... i_8)} &= |\langle \Psi_{ex} | \hat{\sigma}_y^{(i_1)} \otimes ... \otimes \hat{\sigma}_y^{(i_8)}| \Psi_{ex}^* \rangle|^2 \; ,
\label{eq:n-tangle}
\end{align}
where $\sigma_y^{(i_k)}$ is the Pauli matrix acting on visible node $i_k$. 
Such $8$-tangles are directly connected to $4$-nucleon entanglement, and have been computed for $sd$-shell nuclei in Ref.~\cite{Brokemeier:2024lhq}.
Fig.~\ref{fig:tau8_fid} shows the fidelity of the nuclear NQS with $\alpha=1$ as a function of total $8$-tangles in the mixed proton-neutron sector, pure neutron and pure proton sector, defined as
\begin{align}
& \overline{\tau}^{(8)}_{\pi\nu}  \equiv \sum_{\substack{ i_1, i_2, .. i_8 \\ \text{mixed } } } \tau^{(8)}_{(i_1 i_2 .. i_8)} \; , \nonumber \\
\overline{\tau}^{(8)}_{\pi}  \equiv \sum_{\substack{ i_1, i_2, .. i_8 \\ \text{all protons} } }  &\tau^{(8)}_{(i_1 i_2 .. i_8)} \; , \ \ 
 \overline{\tau}^{(8)}_{\nu}  \equiv \sum_{\substack{ i_1, i_2, .. i_8 \\ \text{all neutrons} } } \tau^{(8)}_{(i_1 i_2 .. i_8)} 
\; .
\label{eq:ntangles_sum}
\end{align}

It is clear from Fig.~\ref{fig:tau8_fid} that the NQS accuracy is sensitive to multi-proton-neutron entanglement, in a similar way as the non-stabilizerness. On the other hand we do not observe such sensitivity in the pure proton or pure neutron sectors.
This highlights the role of the proton-neutron interaction in generating quantum complexity.
\\

\begin{figure}
\centering
  \includegraphics[width=\textwidth]{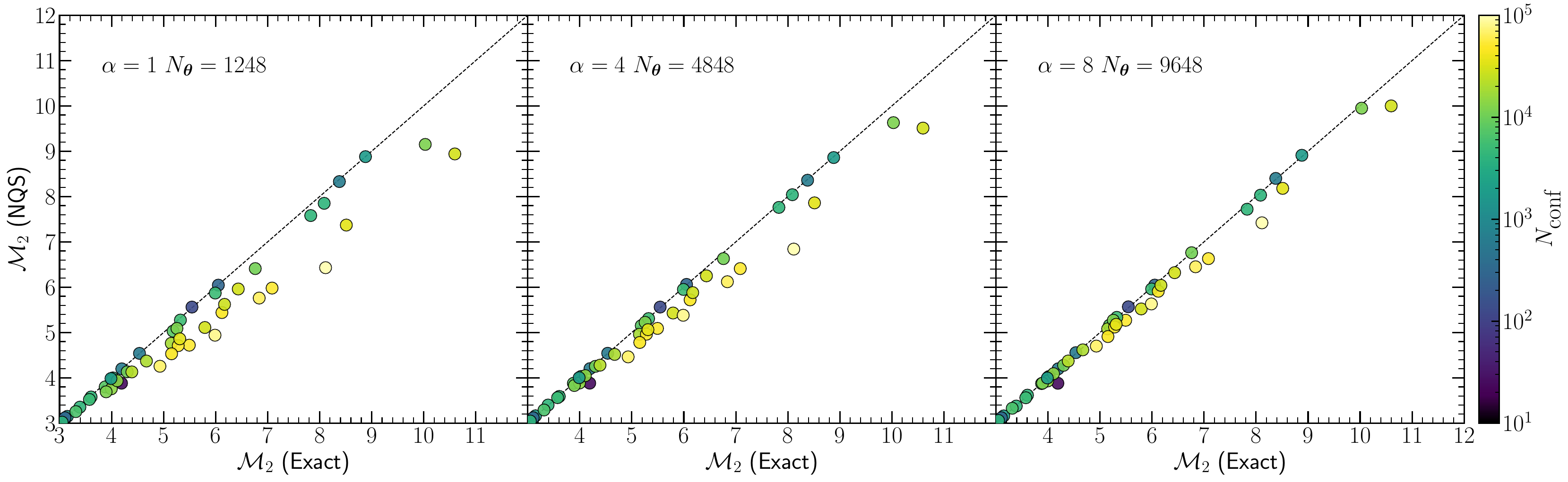}
  \includegraphics[width=\textwidth]{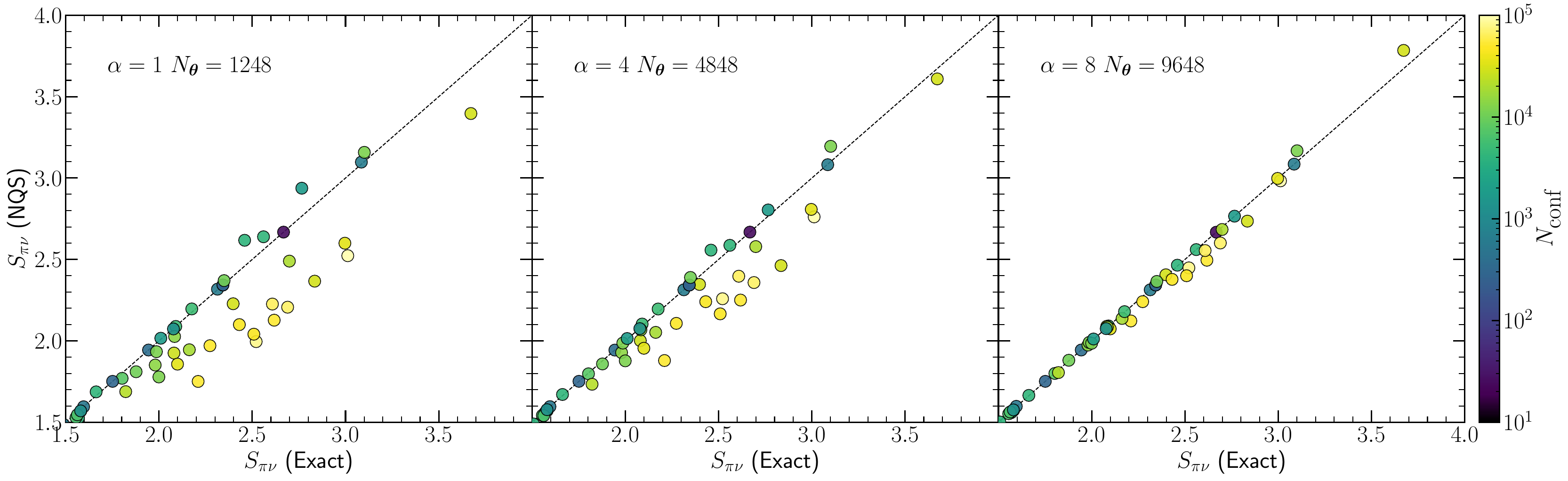}
\caption{Non-stabilizerness and bi-partite proton-neutron entanglement of the nuclear NQS versus exact values, for $\alpha=1,2,8$.}
\label{fig:M2_Spn_ex_vs_NQS}
\end{figure}

\textit{Quantum Complexity of the Network.}--- 
Finally we present in Fig.~\ref{fig:M2_Spn_ex_vs_NQS} measures of non-stabilizerness and entanglement of the nuclear NQS, in comparison with those of the exact state. 
The non-stabilizerness is characterized by the stabilizer 2-R\'enyi entropy $\mathcal{M}_2$, as in the main text, while the entanglement is characterized by bi-partite proton-neutron von Neumann entropy $S_{\pi\nu}$ obtained by tracing over, for example, the neutron sector
\begin{align}
    S_{\pi\nu} = - \text{Tr} \left( \rho_\pi \, \log_2 \rho_\pi \right) \; , \ \ \rho_\pi = \text{Tr}_{\nu} (\ket{\Psi} \bra{\Psi}) \; .
\end{align}
This measure is simpler to compute that the $n$-tangles and represents, in some way, the residue of proton-neutron multipartite entanglement.
Only nuclei with $\mathcal{M}_2>3$ or $S_{\pi\nu} > 1.5$ are displayed, as those with lower values of quantum complexity are reproduced to great accuracy.
It is seen that the network systematically underestimates the non-stabilizer content of the exact state, while entanglement can be overestimated for some nuclei.

\section*{Supplemental Material}
We show in Table~\ref{tab:fidelity_full} the fidelity $\mathcal{F}_{\boldsymbol{\theta}}$ of the fidelity-maximized RBM with respect to the exact ground state $\ket{\Psi_{ex}}$, for $\alpha=1,2,4,8$, as plotted in Fig.~\ref{fig:fidmax_M2}. 
Tables~\ref{tab:VMC_full_1} and \ref{tab:VMC_full_2} show the relative energy error and fidelity of the VMC-energy-minimized RBM, as plotted in Fig.~\ref{fig:VMC_M2}. 
In order to appreciate how the network performs depending on the non-stabilizer content of the nuclear state, nuclei are organized by configuration numbers $N_\mathrm{conf}$, and increasing values of $\mathcal{M}_2(\ket{\Psi_{ex}})$ for each $N_\mathrm{conf}$.
\begin{table*}
\centering
\small
\begin{minipage}[t]{0.495\textwidth}\vspace{0pt}
\centering
\renewcommand{\arraystretch}{1.2}
\begin{tabular}{c c c c c c c}
\hline
\hline
$N_{\mathrm{conf}}$ & Nucl. & $\mathcal{M}_2(|\Psi_{ex}\rangle)$ & \multicolumn{4}{c}{Fidelity $\mathcal{F}_{\boldsymbol{\theta}}$} \\
&&& $\alpha = 1$ & $\alpha = 2$ & $\alpha = 4$ & $\alpha = 8$ \\
\hline
\hline
10 & $^{38}$K & 0.081 & 1.000 & 1.000 & 1.000 & 1.000 \\
 & $^{28}$F & 0.107 & 1.000 & 1.000 & 1.000 & 1.000 \\
\hline
14 & $^{26}$O & 0.309 & 1.000 & 1.000 & 1.000 & 1.000 \\
 & $^{38}$Ar & 0.312 & 1.000 & 1.000 & 1.000 & 1.000 \\
 & $^{30}$Ne & 1.200 & 1.000 & 1.000 & 1.000 & 1.000 \\
 & $^{18}$O & 1.774 & 1.000 & 1.000 & 1.000 & 1.000 \\
\hline
22 & $^{31}$Na & 0.395 & 1.000 & 1.000 & 1.000 & 1.000 \\
 & $^{19}$O & 0.968 & 1.000 & 1.000 & 1.000 & 1.000 \\
\hline
25 & $^{18}$F & 4.191 & 1.000 & 1.000 & 1.000 & 1.000 \\
\hline
32 & $^{37}$Cl & 0.284 & 1.000 & 1.000 & 1.000 & 1.000 \\
 & $^{25}$O & 0.289 & 1.000 & 1.000 & 1.000 & 1.000 \\
\hline
79 & $^{27}$F & 1.199 & 1.000 & 1.000 & 1.000 & 1.000 \\
\hline
80 & $^{33}$Al & 0.520 & 1.000 & 1.000 & 1.000 & 1.000 \\
 & $^{21}$O & 1.210 & 1.000 & 1.000 & 1.000 & 1.000 \\
\hline
81 & $^{24}$O & 0.524 & 1.000 & 1.000 & 1.000 & 1.000 \\
 & $^{36}$S & 0.533 & 1.000 & 1.000 & 1.000 & 1.000 \\
 & $^{32}$Mg & 1.498 & 1.000 & 1.000 & 1.000 & 1.000 \\
 & $^{20}$O & 1.940 & 1.000 & 1.000 & 1.000 & 1.000 \\
\hline
109 & $^{37}$Ar & 1.804 & 1.000 & 1.000 & 1.000 & 1.000 \\
 & $^{29}$Ne & 1.961 & 1.000 & 1.000 & 1.000 & 1.000 \\
\hline
119 & $^{35}$P & 0.472 & 1.000 & 1.000 & 1.000 & 1.000 \\
 & $^{23}$O & 0.702 & 1.000 & 1.000 & 1.000 & 1.000 \\
\hline
128 & $^{19}$F & 5.545 & 1.000 & 1.000 & 1.000 & 1.000 \\
\hline
142 & $^{34}$Si & 0.514 & 1.000 & 1.000 & 1.000 & 1.000 \\
 & $^{22}$O & 1.425 & 1.000 & 1.000 & 1.000 & 1.000 \\
\hline
303 & $^{36}$Cl & 0.920 & 1.000 & 1.000 & 1.000 & 1.000 \\
& $^{30}$Na & 3.105 & 1.000 & 1.000 & 1.000 & 1.000 \\
 & $^{20}$F & 6.052 & 1.000 & 1.000 & 1.000 & 1.000 \\
\hline
369 & $^{26}$F & 1.811 & 1.000 & 1.000 & 1.000 & 1.000 \\
\hline
526 & $^{22}$F & 2.552 & 1.000 & 1.000 & 1.000 & 1.000 \\
\hline
579 & $^{25}$F & 1.252 & 1.000 & 1.000 & 1.000 & 1.000 \\
 & $^{21}$F & 4.536 & 0.999 & 0.999 & 1.000 & 1.000 \\
\hline
640 & $^{28}$Ne & 2.733 & 0.999 & 1.000 & 1.000 & 1.000 \\
 & $^{36}$Ar & 4.196 & 0.998 & 1.000 & 1.000 & 1.000 \\
 & $^{20}$Ne & 8.379 & 0.997 & 0.999 & 1.000 & 1.000 \\
\hline
732 & $^{35}$S & 1.174 & 0.998 & 1.000 & 1.000 & 1.000 \\
 & $^{31}$Mg & 2.913 & 0.999 & 1.000 & 1.000 & 0.999 \\
\hline
787 & $^{24}$F & 1.702 & 0.999 & 1.000 & 1.000 & 1.000 \\
\hline
1068 & $^{23}$F & 2.830 & 0.997 & 1.000 & 1.000 & 1.000 \\
\hline
1222 & $^{34}$P & 1.815 & 0.997 & 0.999 & 0.999 & 0.999 \\
 & $^{32}$Al & 2.375 & 0.998 & 0.999 & 0.999 & 0.999 \\
\hline
1322 & $^{33}$Si & 1.371 & 0.998 & 0.999 & 1.000 & 0.999 \\
\hline
\hline
\end{tabular}
\end{minipage}
\hfill
\begin{minipage}[t]{0.495\textwidth}\vspace{0pt}
\centering
\renewcommand{\arraystretch}{1.2}
\begin{tabular}{c c c c c c c}
\hline
\hline
$N_{\mathrm{conf}}$ & Nucl. & $\mathcal{M}_2(|\Psi_{ex}\rangle)$ & \multicolumn{4}{c}{Fidelity $\mathcal{F}_{\boldsymbol{\theta}}$} \\
&&& $\alpha = 1$ & $\alpha = 2$ & $\alpha = 4$ & $\alpha = 8$ \\
\hline
\hline
1395 & $^{29}$Na & 2.651 & 0.995 & 0.999 & 1.000 & 1.000 \\
\hline
1736 & $^{27}$Ne & 3.149 & 0.994 & 0.999 & 1.000 & 0.999 \\
 & $^{35}$Cl & 3.988 & 0.983 & 0.998 & 0.999 & 0.999 \\
 & $^{21}$Ne & 8.881 & 0.922 & 0.990 & 0.999 & 1.000 \\
\hline
3968 & $^{34}$Cl & 3.049 & 0.982 & 0.993 & 0.997 & 0.999 \\
 & $^{22}$Na & 8.087 & 0.868 & 0.952 & 0.990 & 0.998 \\
 \hline
4206 & $^{34}$S & 3.391 & 0.964 & 0.988 & 0.993 & 0.995 \\
 & $^{26}$Ne & 3.606 & 0.980 & 0.991 & 0.993 & 0.997 \\
 & $^{30}$Mg & 4.025 & 0.973 & 0.992 & 0.998 & 0.997 \\
 & $^{22}$Ne & 7.828 & 0.872 & 0.954 & 0.989 & 0.993 \\
\hline
4882 & $^{31}$Al & 3.574 & 0.977 & 0.992 & 0.998 & 0.998 \\
 & $^{23}$Ne & 5.993 & 0.932 & 0.978 & 0.994 & 0.997 \\
\hline
5052 & $^{28}$Na & 5.323 & 0.963 & 0.987 & 0.996 & 0.998 \\
\hline
6457 & $^{33}$P & 3.313 & 0.963 & 0.985 & 0.993 & 0.997 \\
 & $^{25}$Ne & 4.011 & 0.964 & 0.987 & 0.993 & 0.996 \\
\hline
7562 & $^{32}$Si & 3.877 & 0.959 & 0.982 & 0.994 & 0.993 \\
 & $^{24}$Ne & 5.182 & 0.933 & 0.973 & 0.990 & 0.993 \\
\hline
10014 & $^{27}$Na & 4.300 & 0.937 & 0.969 & 0.989 & 0.996 \\
\hline
10435 & $^{24}$Na & 6.762 & 0.867 & 0.931 & 0.973 & 0.991 \\
\hline
11940 & $^{33}$S & 3.896 & 0.928 & 0.961 & 0.983 & 0.993 \\
 & $^{29}$Mg & 5.256 & 0.930 & 0.966 & 0.987 & 0.995 \\
 & $^{23}$Na & 10.030 & 0.707 & 0.829 & 0.918 & 0.972 \\
\hline
14099 & $^{26}$Na & 3.995 & 0.929 & 0.961 & 0.983 & 0.993 \\
 & $^{30}$Al & 4.107 & 0.937 & 0.965 & 0.984 & 0.992 \\
\hline
18411 & $^{25}$Na & 5.147 & 0.876 & 0.928 & 0.965 & 0.985 \\
\hline
19728 & $^{32}$P & 4.670 & 0.905 & 0.944 & 0.970 & 0.986 \\
\hline
21713 & $^{31}$Si & 4.389 & 0.919 & 0.954 & 0.976 & 0.989 \\
\hline
26914 & $^{26}$Al & 6.171 & 0.845 & 0.903 & 0.946 & 0.975 \\
\hline
28503 & $^{32}$S & 5.793 & 0.813 & 0.883 & 0.928 & 0.952 \\
 & $^{28}$Mg & 6.435 & 0.846 & 0.909 & 0.953 & 0.979 \\
 & $^{24}$Mg & 10.598 & 0.585 & 0.698 & 0.806 & 0.895 \\
\hline
34971 & $^{29}$Al & 5.311 & 0.877 & 0.924 & 0.957 & 0.978 \\
 & $^{25}$Mg & 8.512 & 0.718 & 0.806 & 0.878 & 0.936 \\
\hline
44133 & $^{31}$P & 5.153 & 0.846 & 0.895 & 0.932 & 0.956 \\
 & $^{27}$Mg & 6.121 & 0.825 & 0.888 & 0.934 & 0.963 \\
\hline
49884 & $^{28}$Al & 5.281 & 0.853 & 0.902 & 0.941 & 0.967 \\
\hline
51630 & $^{30}$Si & 5.494 & 0.811 & 0.866 & 0.924 & 0.956 \\
 & $^{26}$Mg & 7.085 & 0.738 & 0.818 & 0.886 & 0.929 \\
\hline
64299 & $^{27}$Al & 6.838 & 0.754 & 0.825 & 0.876 & 0.932 \\
\hline
67251 & $^{30}$P & 4.930 & 0.849 & 0.886 & 0.922 & 0.961 \\
\hline
80115 & $^{29}$Si & 5.988 & 0.760 & 0.829 & 0.894 & 0.935 \\
\hline
93710 & $^{28}$Si & 8.113 & 0.640 & 0.730 & 0.789 & 0.872 \\
\hline
\hline
\end{tabular}
\end{minipage}
\caption{Fidelity $\mathcal{F}_{\boldsymbol{\theta}}$ of the fidelity-maximized RBM for different values of $\alpha$. The nuclei are organized by configuration numbers $N_\mathrm{conf}$, and increasing $\mathcal{M}_2(\ket{\Psi_{ex}})$ values. The number of network parameters for $\alpha=1,2,4,8$ is $N_{\btheta}= 1248, \, 2448, \, 4848, \, 9648$, respectively.}
\label{tab:fidelity_full}
\end{table*}

\begin{table*}
\centering
\small
\begin{minipage}[t]{\textwidth}\vspace{0pt}
\centering
\renewcommand{\arraystretch}{1.2}
\begin{tabular}{c c c c c c c c c}
\hline
\hline
 $N_{\mathrm{conf}}$ & Nucl. & $\mathcal{M}_2(|\Psi_{ex}\rangle)$ & \multicolumn{3}{c}{Energy error $\varepsilon_{\boldsymbol{\theta}}$} &  \multicolumn{3}{c}{Fidelity $\mathcal{F}_{\boldsymbol{\theta}}$}\\ 
&&& $\alpha = 2$ & $\alpha = 4$ & $\alpha = 8$ & $\alpha = 2$ & $\alpha = 4$ & $\alpha = 8$\\
\hline
\hline
10 & $^{38}$K & 0.081 & $8.31 \times 10^{-10}$ & $6.55 \times 10^{-9}$ & $1.82 \times 10^{-9}$ & 1.000 & 1.000 & 1.000 \\
 & $^{28}$F & 0.107 & $2.73 \times 10^{-13}$ & $8.09 \times 10^{-14}$ & $4.08 \times 10^{-13}$ & 1.000 & 1.000 & 1.000 \\
\hline
14 & $^{26}$O & 0.309 & $4.42 \times 10^{-10}$ & $6.97 \times 10^{-12}$ & $1.08 \times 10^{-8}$ & 1.000 & 1.000 & 1.000 \\
 & $^{38}$Ar & 0.312 & $1.56 \times 10^{-10}$ & $3.93 \times 10^{-11}$ & $1.66 \times 10^{-11}$ & 1.000 & 1.000 & 1.000 \\
 & $^{30}$Ne & 1.200 & $1.85 \times 10^{-11}$ & $1.93 \times 10^{-11}$ & $3.73 \times 10^{-10}$ & 1.000 & 1.000 & 1.000 \\
 & $^{18}$O & 1.774 & $6.45 \times 10^{-10}$ & $1.18 \times 10^{-9}$ & $5.33 \times 10^{-11}$ & 1.000 & 1.000 & 1.000 \\
\hline
22 & $^{31}$Na & 0.395 & $3.51 \times 10^{-12}$ & $1.34 \times 10^{-12}$ & $5.05 \times 10^{-11}$ & 1.000 & 1.000 & 1.000 \\
 & $^{19}$O & 0.968 & $3.20 \times 10^{-11}$ & $4.98 \times 10^{-12}$ & $4.95 \times 10^{-10}$ & 1.000 & 1.000 & 1.000 \\
\hline
25 & $^{18}$F & 4.191 & $6.29 \times 10^{-10}$ & $2.55 \times 10^{-9}$ & $7.55 \times 10^{-10}$ & 1.000 & 1.000 & 1.000 \\
\hline
32 & $^{37}$Cl & 0.284 & $3.37 \times 10^{-10}$ & $1.85 \times 10^{-10}$ & $1.60 \times 10^{-9}$ & 1.000 & 1.000 & 1.000 \\
 & $^{25}$O & 0.289 & $2.37 \times 10^{-7}$ & $5.15 \times 10^{-11}$ & $3.57 \times 10^{-9}$ & 1.000 & 1.000 & 1.000 \\
\hline
79 & $^{27}$F & 1.199 & $5.07 \times 10^{-6}$ & $4.35 \times 10^{-10}$ & $1.59 \times 10^{-7}$ & 1.000 & 1.000 & 1.000 \\
\hline
80 & $^{33}$Al & 0.520 & $3.73 \times 10^{-4}$ & $1.41 \times 10^{-4}$ & $3.84 \times 10^{-7}$ & 0.996 & 0.998 & 1.000 \\
 & $^{21}$O & 1.210 & $9.90 \times 10^{-7}$ & $1.35 \times 10^{-6}$ & $1.83 \times 10^{-8}$ & 1.000 & 1.000 & 1.000 \\
\hline
81 & $^{24}$O & 0.524 & $5.12 \times 10^{-8}$ & $2.90 \times 10^{-8}$ & $6.28 \times 10^{-6}$ & 1.000 & 1.000 & 1.000 \\
 & $^{36}$S & 0.533 & $9.48 \times 10^{-6}$ & $6.40 \times 10^{-10}$ & $3.09 \times 10^{-9}$ & 1.000 & 1.000 & 1.000 \\
 & $^{32}$Mg & 1.498 & $6.08 \times 10^{-7}$ & $1.34 \times 10^{-6}$ & $3.35 \times 10^{-10}$ & 1.000 & 1.000 & 1.000 \\
 & $^{20}$O & 1.940 & $2.32 \times 10^{-10}$ & $2.82 \times 10^{-9}$ & $7.43 \times 10^{-10}$ & 1.000 & 1.000 & 1.000 \\
\hline
109 & $^{37}$Ar & 1.804 & $6.44 \times 10^{-5}$ & $7.13 \times 10^{-11}$ & $1.973\times 10^{-11}$ & 0.999 & 1.000 & 1.000 \\
 & $^{29}$Ne & 1.961 & $3.02 \times 10^{-6}$ & $1.84 \times 10^{-8}$ & $2.43 \times 10^{-10}$ & 1.000 & 1.000 & 1.000 \\
\hline
119 & $^{35}$P & 0.472 & $2.91 \times 10^{-8}$ & $1.77 \times 10^{-4}$ & $2.26 \times 10^{-8}$ & 1.000 & 0.998 & 1.000 \\
 & $^{23}$O & 0.702 & $1.62 \times 10^{-6}$ & $1.22 \times 10^{-7}$ & $5.37 \times 10^{-7}$ & 1.000 & 1.000 & 1.000 \\
\hline
128 & $^{19}$F & 5.545 & $3.03 \times 10^{-4}$ & $8.77 \times 10^{-4}$ & $4.38 \times 10^{-5}$ & 0.988 & 0.730 & 0.985 \\
\hline
142 & $^{34}$Si & 0.514 & $7.81 \times 10^{-5}$ & $4.19 \times 10^{-8}$ & $1.46 \times 10^{-6}$ & 0.999 & 1.000 & 1.000 \\
 & $^{22}$O & 1.425 & $1.40 \times 10^{-3}$ & $9.03 \times 10^{-4}$ & $8.00 \times 10^{-8}$ & 0.997 & 0.997 & 1.000 \\
\hline
303 & $^{36}$Cl & 0.920 & $2.01 \times 10^{-7}$ & $1.55 \times 10^{-7}$ & $3.12 \time 10^{-07}$$ $ & 1.000 & 1.000 & 1.000 \\
 & $^{30}$Na & 3.105 & $3.16 \times 10^{-5}$ & $3.37 \times 10^{-7}$ & $4.93 \times 10^{-6}$ & 0.999 & 1.000 & 0.999 \\
 & $^{20}$F & 6.052 & $1.04 \times 10^{-5}$ & $1.52 \times 10^{-6}$ & $1.19 \times 10^{-6}$ & 1.000 & 1.000 & 1.000 \\
\hline
369 & $^{26}$F & 1.811 & $5.67 \times 10^{-5}$ & $1.20 \times 10^{-5}$ & $8.72 \times 10^{-5}$ & 0.998 & 1.000 & 0.999 \\
\hline
526 & $^{22}$F & 2.552 & $3.70 \times 10^{-4}$ & $1.30 \times 10^{-5}$ & $5.09 \times 10^{-6}$ & 0.999 & 1.000 & 1.000 \\
\hline
579 & $^{25}$F & 1.252 & $1.77 \times 10^{-4}$ & $1.11 \times 10^{-6}$ & $5.32 \times 10^{-5}$ & 0.999 & 1.000 & 1.000 \\
 & $^{21}$F & 4.536 & $3.33 \times 10^{-5}$ & $2.10 \times 10^{-6}$ & $6.81 \times 10^{-6}$ & 1.000 & 1.000 & 1.000 \\
\hline
640 & $^{28}$Ne & 2.733 & $1.57 \times 10^{-3}$ & $1.19 \times 10^{-5}$ & $3.27 \times 10^{-5}$ & 0.986 & 1.000 & 1.000 \\
 & $^{36}$Ar & 4.196 & $2.53 \times 10^{-5}$ & $7.03 \times 10^{-6}$ & $2.94 \times 10^{-4}$ & 0.999 & 1.000 & 0.996 \\
 & $^{20}$Ne & 8.379 & $7.76 \times 10^{-7}$ & $2.64 \times 10^{-8}$ & $2.40 \times 10^{-7}$ & 1.000 & 1.000 & 1.000 \\
\hline
732 & $^{35}$S & 1.174 & $6.85 \times 10^{-5}$ & $2.74 \times 10^{-6}$ & $4.62 \times 10^{-5}$ & 0.999 & 1.000 & 0.999 \\
 & $^{31}$Mg & 2.913 & $6.29 \times 10^{-5}$ & $1.20 \times 10^{-6}$ & $3.36 \times 10^{-5}$ & 1.000 & 1.000 & 0.999 \\
\hline
787 & $^{24}$F & 1.702 & $1.78 \times 10^{-3}$ & $1.31 \times 10^{-5}$ & $7.01 \times 10^{-5}$ & 0.993 & 1.000 & 1.000 \\
\hline
1068 & $^{23}$F & 2.830 & $5.93 \times 10^{-4}$ & $1.49 \times 10^{-4}$ & $4.89 \times 10^{-5}$ & 0.998 & 1.000 & 1.000 \\
\hline
1222 & $^{34}$P & 1.815 & $1.96 \times 10^{-4}$ & $9.36 \times 10^{-5}$ & $1.25 \times 10^{-4}$ & 0.997 & 0.998 & 0.998 \\
 & $^{32}$Al & 2.375 & $1.88 \times 10^{-4}$ & $1.16 \times 10^{-4}$ & $1.43 \times 10^{-4}$ & 0.998 & 0.998 & 0.997 \\
\hline
1322 & $^{33}$Si & 1.371 & $2.26 \times 10^{-4}$ & $6.47 \times 10^{-6}$ & $7.51 \times 10^{-5}$ & 0.998 & 0.999 & 0.999 \\
\hline
\hline
\end{tabular}
\end{minipage}
\caption{Relative energy error $\varepsilon_{\btheta}$ and fidelity $\mathcal{F}_{\boldsymbol{\theta}}$ of the VMC-energy-minimized RBM for different values of $\alpha$. The nuclei are organized by configuration numbers $N_\mathrm{conf}$, and increasing $\mathcal{M}_2(\ket{\Psi_{ex}})$ values. The number of network parameters for $\alpha=1,2,4,8$ is $N_{\btheta}= 1248, \, 2448, \, 4848, \, 9648$, respectively. This table shows nuclei with $N_\mathrm{conf} \leq 1322$.}
\label{tab:VMC_full_1}
\end{table*}
\begin{table*}
\centering
\small
\begin{minipage}[t]{\textwidth}\vspace{0pt}
\centering
\renewcommand{\arraystretch}{1.2}
\begin{tabular}{c c c c c c c c c}
\hline
\hline
 $N_{\mathrm{conf}}$ & Nucl. & $\mathcal{M}_2(|\Psi_{ex}\rangle)$ & \multicolumn{3}{c}{Energy error $\varepsilon_{\boldsymbol{\theta}}$} &  \multicolumn{3}{c}{Fidelity $\mathcal{F}_{\boldsymbol{\theta}}$}\\ 
&&& $\alpha = 2$ & $\alpha = 4$ & $\alpha = 8$ & $\alpha = 2$ & $\alpha = 4$ & $\alpha = 8$\\
\hline
\hline
1395 & $^{29}$Na & 2.651 & $2.54 \times 10^{-4}$ & $5.53 \times 10^{-5}$ & $5.17 \times 10^{-5}$ & 0.998 & 0.999 & 1.000 \\
\hline
1736 & $^{27}$Ne & 3.149 & $3.74 \times 10^{-4}$ & $1.58 \times 10^{-4}$ & $3.49 \times 10^{-4}$ & 0.998 & 0.999 & 0.998 \\
 & $^{35}$Cl & 3.988 & $3.02 \times 10^{-4}$ & $6.02 \times 10^{-5}$ & $1.49 \times 10^{-3}$ & 0.995 & 0.999 & 0.952 \\
 & $^{21}$Ne & 8.881 & $5.23 \times 10^{-3}$ & $4.63 \times 10^{-3}$ & $3.03 \times 10^{-4}$ & 0.744 & 0.285 & 0.977 \\
\hline
3968 & $^{34}$Cl & 3.049 & $5.74 \times 10^{-4}$ & $3.85 \times 10^{-4}$ & $4.71 \times 10^{-4}$ & 0.986 & 0.994 & 0.992 \\
 & $^{22}$Na & 8.087 & $2.42 \times 10^{-2}$ & $4.48 \times 10^{-3}$ & $2.53 \times 10^{-3}$ & 0.625 & 0.925 & 0.953 \\
\hline
4206 & $^{34}$S & 3.391 & $1.55 \times 10^{-3}$ & $9.85 \times 10^{-4}$ & $3.14 \times 10^{-4}$ & 0.977 & 0.988 & 0.996 \\
 & $^{26}$Ne & 3.606 & $2.29 \times 10^{-3}$ & $1.16 \times 10^{-3}$ & $9.42 \times 10^{-4}$ & 0.989 & 0.995 & 0.995 \\
 & $^{30}$Mg & 4.025 & $1.40 \times 10^{-3}$ & $7.42 \times 10^{-4}$ & $9.78 \times 10^{-4}$ & 0.990 & 0.995 & 0.992 \\
 & $^{22}$Ne & 7.828 & $1.97 \times 10^{-2}$ & $6.24 \times 10^{-3}$ & $6.63 \times 10^{-3}$ & 0.823 & 0.976 & 0.965 \\
\hline
4882 & $^{31}$Al & 3.574 & $1.22 \times 10^{-3}$ & $5.62 \times 10^{-4}$ & $3.86 \times 10^{-4}$ & 0.990 & 0.995 & 0.997 \\
 & $^{23}$Ne & 5.993 & $7.51 \times 10^{-3}$ & $2.99 \times 10^{-3}$ & $1.93 \times 10^{-3}$ & 0.952 & 0.989 & 0.993 \\
\hline
5052 & $^{28}$Na & 5.323 & $2.47 \times 10^{-3}$ & $1.35 \times 10^{-3}$ & $7.67 \times 10^{-4}$ & 0.960 & 0.990 & 0.984 \\
\hline
6457 & $^{33}$P & 3.313 & $1.55 \times 10^{-3}$ & $1.17 \times 10^{-3}$ & $7.60 \times 10^{-4}$ & 0.977 & 0.986 & 0.992 \\
 & $^{25}$Ne & 4.011 & $5.05 \times 10^{-3}$ & $2.56 \times 10^{-3}$ & $1.00 \times 10^{-3}$ & 0.974 & 0.989 & 0.994 \\
\hline
7562 & $^{32}$Si & 3.877 & $2.02 \times 10^{-3}$ & $1.16 \times 10^{-3}$ & $6.06 \times 10^{-4}$ & 0.981 & 0.990 & 0.995 \\
 & $^{24}$Ne & 5.182 & $8.74 \times 10^{-3}$ & $3.81 \times 10^{-3}$ & $3.34 \times 10^{-3}$ & 0.961 & 0.985 & 0.988 \\
\hline
10014 & $^{27}$Na & 4.300 & $5.69 \times 10^{-3}$ & $3.18 \times 10^{-3}$ & $1.77 \times 10^{-3}$ & 0.958 & 0.979 & 0.989 \\
\hline
10435 & $^{24}$Na & 6.762 & $1.73 \times 10^{-2}$ & $7.60 \times 10^{-3}$ & $3.94 \times 10^{-3}$ & 0.863 & 0.907 & 0.977 \\
\hline
11940 & $^{33}$S & 3.896 & $3.30 \times 10^{-3}$ & $2.32 \times 10^{-3}$ & $1.63 \times 10^{-3}$ & 0.929 & 0.955 & 0.972 \\
 & $^{29}$Mg & 5.256 & $5.30 \times 10^{-3}$ & $3.04 \times 10^{-3}$ & $1.91 \times 10^{-3}$ & 0.948 & 0.969 & 0.982 \\
 & $^{23}$Na & 10.030 & $2.24 \times 10^{-2}$ & $1.64 \times 10^{-2}$ & $1.27 \times 10^{-2}$ & 0.374 & 0.466 & 0.550 \\
\hline
14099 & $^{26}$Na & 3.995 & $6.80 \times 10^{-3}$ & $5.03 \times 10^{-3}$ & $3.91 \times 10^{-3}$ & 0.949 & 0.967 & 0.978 \\
 & $^{30}$Al & 4.107 & $4.24 \times 10^{-3}$ & $2.69 \times 10^{-3}$ & $6.91 \times 10^{-3}$ & 0.926 & 0.949 & 0.709 \\
\hline
18411 & $^{25}$Na & 5.147 & $1.36 \times 10^{-2}$ & $8.75 \times 10^{-3}$ & $5.07 \times 10^{-3}$ & 0.895 & 0.941 & 0.968 \\
\hline
19728 & $^{32}$P & 4.670 & $4.34 \times 10^{-3}$ & $3.04 \times 10^{-3}$ & $2.41 \times 10^{-3}$ & 0.737 & 0.278 & 0.358 \\
\hline
21713 & $^{31}$Si & 4.389 & $4.86 \times 10^{-3}$ & $3.27 \times 10^{-3}$ & $2.23 \times 10^{-3}$ & 0.925 & 0.957 & 0.972 \\
\hline
26914 & $^{26}$Al & 6.171 & $1.43 \times 10^{-2}$ & $9.71 \times 10^{-3}$ & $6.96 \times 10^{-3}$ & 0.853 & 0.911 & 0.941 \\
\hline
28503 & $^{32}$S & 5.793 & $8.19 \times 10^{-3}$ & $6.48 \times 10^{-3}$ & $4.60 \times 10^{-3}$ & 0.781 & 0.836 & 0.887 \\
 & $^{28}$Mg & 6.435 & $1.37 \times 10^{-2}$ & $9.38 \times 10^{-3}$ & $6.94 \times 10^{-3}$ & 0.862 & 0.929 & 0.946 \\
 & $^{24}$Mg & 10.598 & $4.98 \times 10^{-2}$ & $4.04 \times 10^{-2}$ & $2.89 \times 10^{-2}$ & 0.375 & 0.492 & 0.799 \\
\hline
34971 & $^{29}$Al & 5.311 & $1.00 \times 10^{-2}$ & $6.97 \times 10^{-3}$ & $4.75 \times 10^{-3}$ & 0.893 & 0.932 & 0.956 \\
 & $^{25}$Mg & 8.512 & $2.56 \times 10^{-2}$ & $1.89 \times 10^{-2}$ & $1.68 \times 10^{-2}$ & 0.627 & 0.691 & 0.870 \\
\hline
44133 & $^{31}$P & 5.153 & $8.50 \times 10^{-3}$ & $6.58 \times 10^{-3}$ & $5.66 \times 10^{-3}$ & 0.828 & 0.877 & 0.899 \\
 & $^{27}$Mg & 6.121 & $1.64 \times 10^{-2}$ & $1.17 \times 10^{-2}$ & $8.44 \times 10^{-3}$ & 0.825 & 0.879 & 0.924 \\
\hline
49884 & $^{28}$Al & 5.281 & $1.35 \times 10^{-2}$ & $9.79 \times 10^{-3}$ & $5.82 \times 10^{-3}$ & 0.798 & 0.875 & 0.936 \\
\hline
51630 & $^{30}$Si & 5.494 & $1.20 \times 10^{-2}$ & $8.93 \times 10^{-3}$ & $7.12 \times 10^{-3}$ & 0.820 & 0.879 & 0.911 \\
 & $^{26}$Mg & 7.085 & $2.48 \times 10^{-2}$ & $1.89 \times 10^{-2}$ & $1.60 \times 10^{-2}$ & 0.708 & 0.801 & 0.837 \\
\hline
64299 & $^{27}$Al & 6.838 & $1.83 \times 10^{-2}$ & $1.47 \times 10^{-2}$ & $1.16 \times 10^{-2}$ & 0.739 & 0.799 & 0.852 \\
\hline
67251 & $^{30}$P & 4.930 & $1.07 \times 10^{-2}$ & $7.79 \times 10^{-3}$ & $5.95 \times 10^{-3}$ & 0.834 & 0.884 & 0.896 \\
\hline
80115 & $^{29}$Si & 5.988 & $1.39 \times 10^{-2}$ & $1.12 \times 10^{-2}$ & $9.53 \times 10^{-3}$ & 0.772 & 0.827 & 0.865 \\
\hline
93710 & $^{28}$Si & 8.113 & $2.24 \times 10^{-2}$ & $1.91 \times 10^{-2}$ & $1.53 \times 10^{-2}$ & 0.578 & 0.657 & 0.745 \\
\hline
\hline
\end{tabular}
\end{minipage}
\caption{Same as Table~\ref{tab:VMC_full_1} for nuclei with $N_\mathrm{conf} > 1322$.}
\label{tab:VMC_full_2}
\end{table*}

\end{document}